\documentclass[fleqn,12pt,twoside]{article}
\usepackage{espcrc1}

\usepackage{epsfig}

\usepackage[figuresright]{rotating}

\newcommand{\AmS}{{\protect\the\textfont2
  A\kern-.1667em\lower.5ex\hbox{M}\kern-.125emS}}
\newcommand{\lsim}{\,{\buildrel < \over {_\sim}}\,}

\title{On predictions of the first results from RHIC}

\author{Kari J. Eskola\address{Department of Physics, 
University of Jyv\"askyl\"a,\\ 
P.O. Box 35, FIN-40351 Jyv\"askyl\"a, Finland
}}
       
\begin{document}

\maketitle

\begin{abstract}
In this talk, I will discuss the predictions of the first results from
RHIC: the charged particle multiplicity $dN_{\rm ch}/d\eta$, its centrality 
dependence and the elliptic flow $v_2$.
\end{abstract}

\begin{flushleft}
\vspace{-8.5cm}
JYFL-05/01\\
hep-ph/0104058
\vspace{7.5cm}
\end{flushleft}

\section{Charged particle multiplicity in central collisions}

\subsection{Predictions}

To predict particle multiplicities in $pp$ and $AA$ collisions from
first principles has so far been impossible. Consequently, various
models have been introduced. In Fig.~1 I have collected predictions
for the charged particle multiplicity $dN_{\rm ch}/dy$ at $y=0$ for Au-Au
collisions at $\sqrt s=200$ $A$GeV, presented at {\em Quark Matter
'99} \cite{LAST_CALL} and after, until July 2000, i.e. prior to the
release of the very first data from RHIC.  From the top, the figure
shows predictions from event generator models {\em Heavy Ion Jet
InteractioN event Generator} \cite{HIJING_qm99}, HIJING+{\em Zhang's
Parton Cascade}+{\em A Relativistic Transport model} \cite{AMPT_qm99},
{\em Relativistic Quantum Molecular Dynamics} \cite{RQMD_qm99}, {\em
Ultrarelativistic Quantum Molecular Dynamics} \cite{UrQMD_qm99},
VNI+UrQMD \cite{VNI_UrQMD}, {\em Hadron String Dynamics} and VNI+HSD
\cite{HSD_qm99}, NEXUS \cite{NEXUS_qm99}, {\em Dual Parton Model}
\cite{DPM}, DPMJET \cite{DPMJET}, {\em String Fusion Model}
\cite{AP00}, and from other models, such as {\em Linear EXtrapolation
of Ultrarelativistic nucleon-nucleon Scattering to Nucleus-nucleus
collisions} \cite{LEXUS_6/00}, EKRT saturation model \cite{EKRT},
Hydrodynamics+UrQMD \cite{Hydro_UrQMD}, thermal fireball model
\cite{FBALL_qm99} and McLerran-Venugopalan model \cite{McLV_7/00}. The
assumptions of the underlying QCD dynamics, and even the degrees of
freedom (partons, strings, hadrons, classical fields) vary between
different models, resulting in a variation of a factor two in the
predictions.  The multiplicities at $\sqrt s=200$ $A$GeV are yet to be
measured but, based on the PHOBOS data at 56 and 130 $A$GeV
\cite{PHOBOS} and an extrapolation to 200 $A$GeV (the region between
the dotted lines), it seems that most models predict too many
particles per unit rapidity.

\subsection{Comparison with the first data}
Given the large theoretical uncertainties, the appearance of the first
RHIC data on $dN_{\rm ch}/d\eta$ from Au-Au collisions at $\sqrt
s=130$ $A$GeV \cite{PHOBOS} is greeted with a great enthusiasm. The
model predictions can now be compared with actual measurements for the
first time at collider energies.  In Fig.~2, I have shown examples of
such comparisons. In LEXUS \cite{LEXUS}, one first fits all the
parameters at nucleon-nucleon level. Extrapolation to
ultra-relativistic heavy ion collisions is then done by assuming
different dynamics. As nicely illustrated in the figures, the
``linear'' extrapolation to $AA$, where every binary collision
contributes to particle production, gives the best fit at the SPS,
whereas the wounded nucleon model fits best at both RHIC energies. A
suppression mechanism of the produced quanta is also needed in
string-based models, such as in the {\em Dual String Model}
\cite{DSM+F}, where fusion of strings is needed to fit the RHIC
data. HIJING \cite{HIJING_91} combines pQCD parton dynamics with that
of strings. On one hand, a strong gluon shadowing in the hard pQCD
component of HIJING reduces the multiplicities, but on the other hand
this is partly compensated by jet quenching which feeds some of the
jet energy into the system, resulting in the multiplicities close to
the measured values \cite{WG00}. The EKRT saturation model, which is
based on pQCD parton dynamics supplemented by a requirement of
saturation of produced gluons also predicted correctly both the
normalization and the $\sqrt s$ scaling for central collisions, as
shown by the dotted-dashed line in Fig.~2 (lower left).  Another model
where the observed multiplicities can be well explained is AMPT, {\em
A MultiPhase Transport Model} \cite{AMPT}, which takes the initial
conditions for the partonic and hadronic afterburners from HIJING and
fits some of the model parameters from the SPS data.  Of the above
models only \cite{EKRT} is a published prediction for 56 and 130
$A$GeV but to my knowledge also \cite{LEXUS_6/00}, \cite{WG00} (see
Fig.~3) and \cite{AMPT} are predictions in the sense that the model
parameters were not tuned to fit the first RHIC data.

\begin{figure}[thb]
\vspace{-13.2cm}
\flushleft{ \hspace{-0.7cm} \epsfxsize=9.5cm\epsfbox{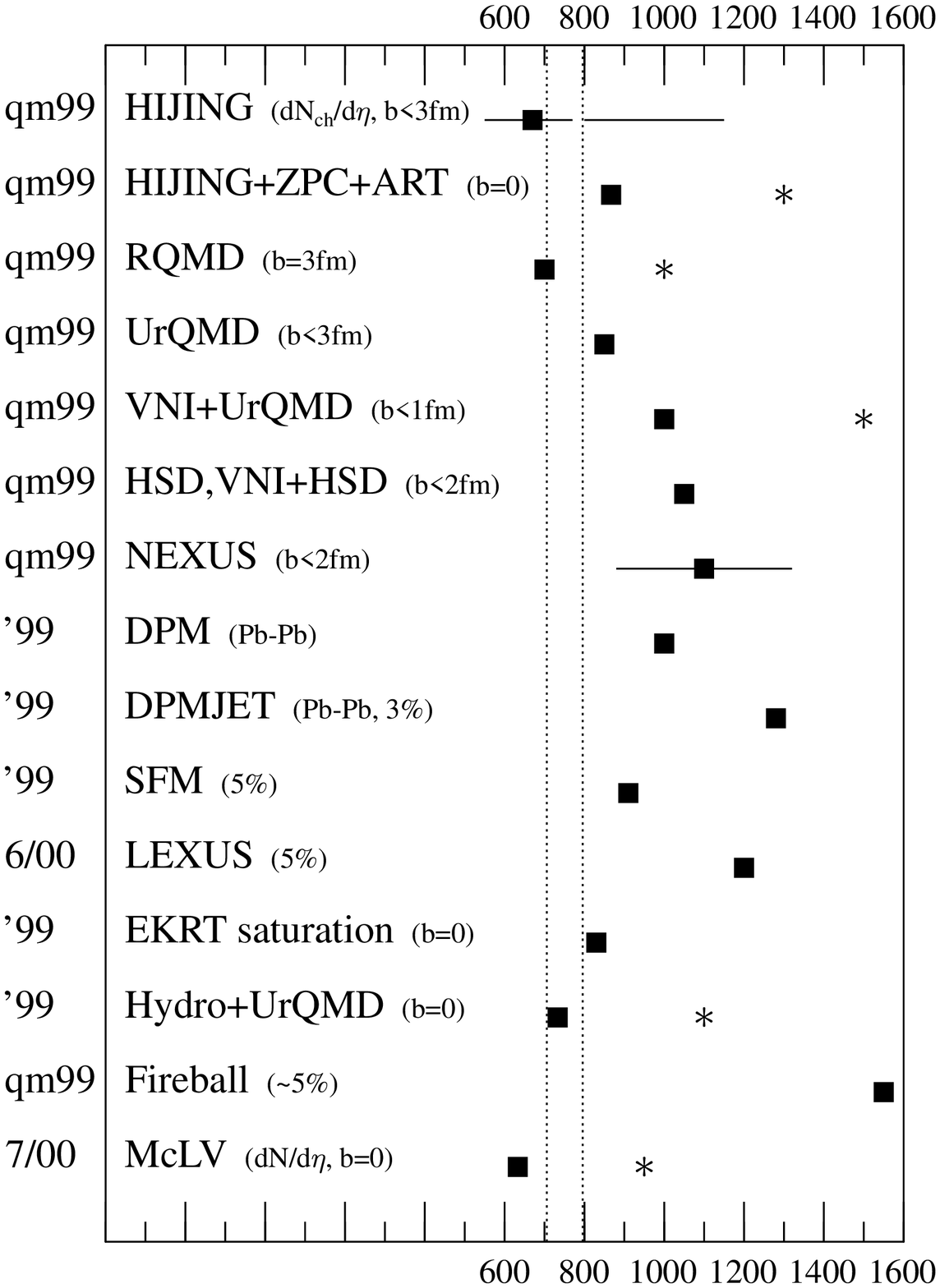}}
\vspace{-15.cm}
\label{qm99}
\end{figure}
\begin{flushright}
\begin{minipage}{6cm}
Figure 1.  {\baselineskip=2pt \small Predictions for $dN_{\rm ch}/dy$
in Au+Au at y=0, $\sqrt s=200$ $A$GeV from different models before the
appearance of the first RHIC data. The information in the parentheses
specifies the parameters used. The time of releasing
the result is indicated on the left.  An approximate factor $2/3$ has
been applied to convert $N_{\rm tot}$ (asterisks) to $N_{\rm ch}$
(boxes). Factors 1.1 and 0.9 approximately accounting for the
conversion of $\eta$ into $y$ and for a $\sim$ 5\% centrality
selection, respectively, have {\em not} been applied in the
figure. The vertical dotted lines are obtained by scaling the PHOBOS data
at 130 $A$GeV
for $dN_{\rm ch}/d\eta$ (6\% central) up by $1.1*(200/130)^{0.37}$ based
on the observed $\sqrt s$-scaling and to account for the conversion of
$\eta$ into $y$. In preparation of this figure I have partly used the
review \cite{AP00}.}
\end{minipage}
\vspace{0.5cm}
\end{flushright}

\vspace{-0.0cm}
An implication from the comparison in Fig.~2 is that coherence
phenomena in particle production become important at RHIC energies:
particle multiplicities are less than what could be expected based on
a mere linear extrapolation from SPS to RHIC based on $pp$ physics.  
It also seems that
efficient final state interactions are needed, which in turn points
towards an early formation of pressure and thermalization. It is also
obvious that even within each model the theoretical uncertainties
related to the fact that the underlying dynamics is not precisely
known in advance, are clearly larger than the experimental
(systematic) errorbars of the data. Therefore, the measured
multiplicities are indispensable constraints for the models. Essential
further constraints, hopefully stringent enough to rule out some
models, will be obtained from the other global observable,
$dE_T/d\eta$, which is more sensitive to the actual evolution of the
system than $dN_{\rm ch}/d\eta$.

\begin{figure}[htb]
\vspace{-0.8cm}
\flushleft{\epsfxsize=8cm\epsfbox{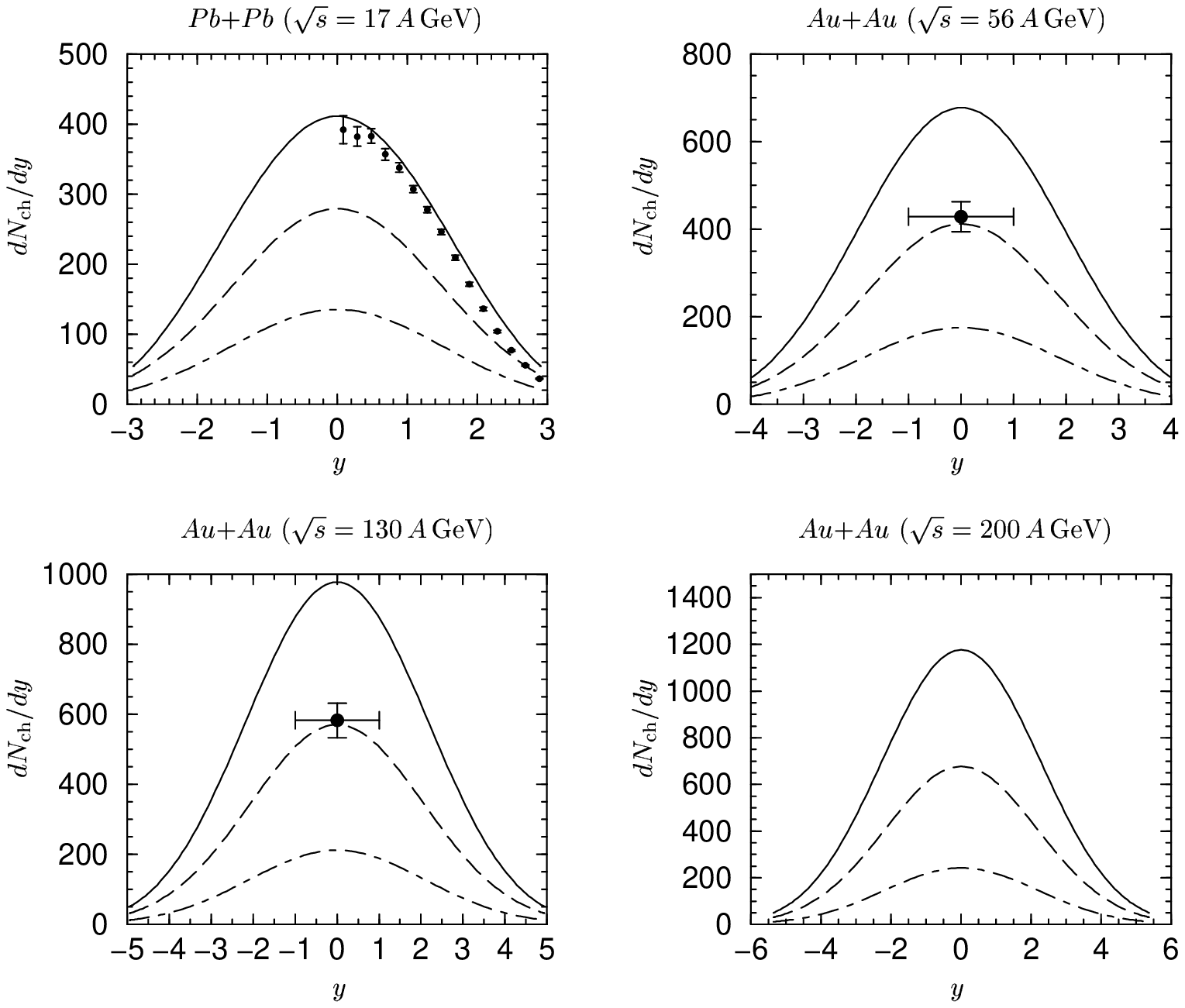}\hspace{0cm}}
\vspace{-7cm}
\flushright{\epsfxsize=7cm\epsfbox{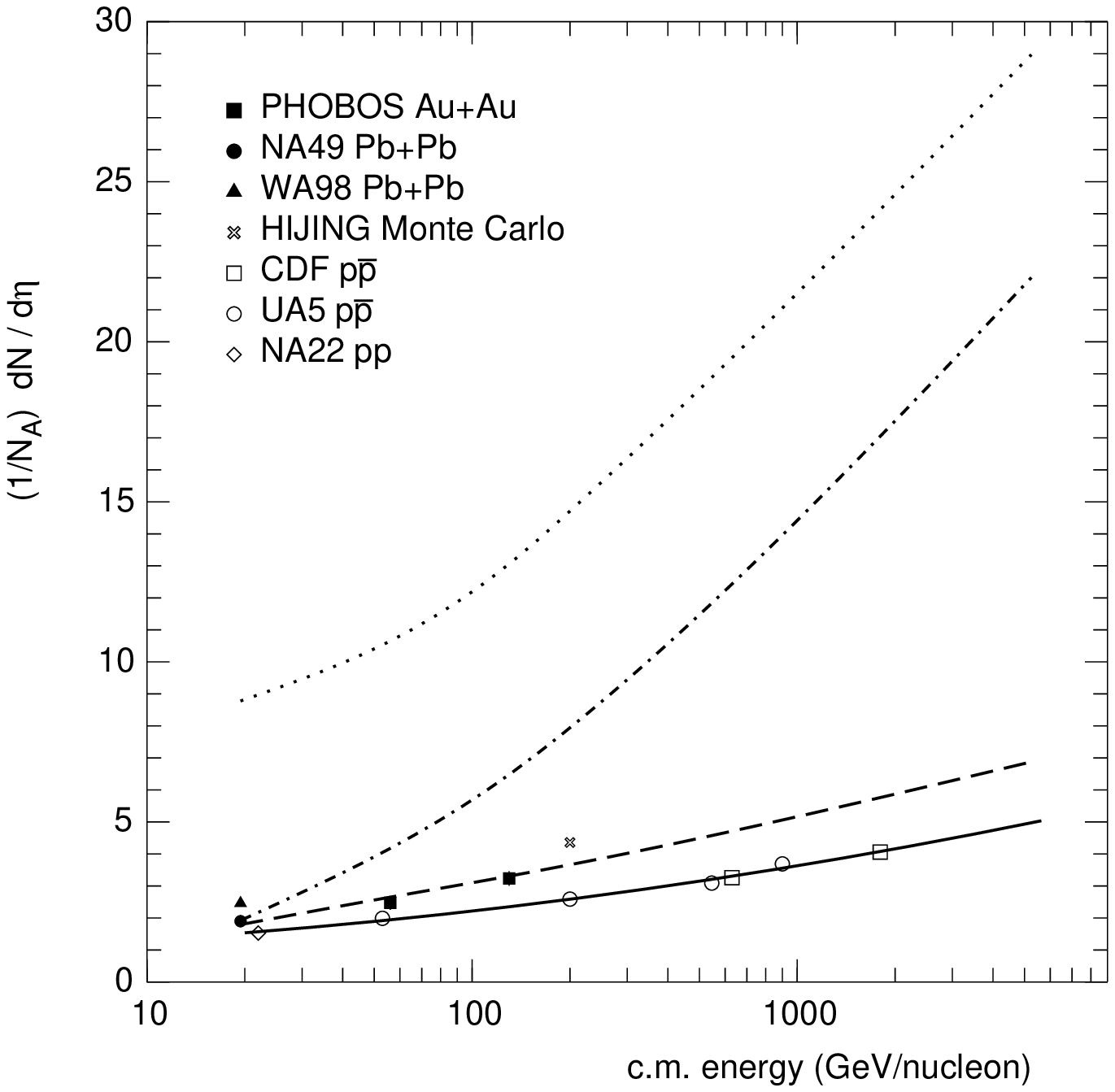}}
\vspace{-2.0cm}
\end{figure}

\setcounter{figure}{1}
\begin{figure}[hbt]
\vspace{0.0cm}
\flushleft{\hspace{1cm}\epsfxsize=6cm\epsfbox{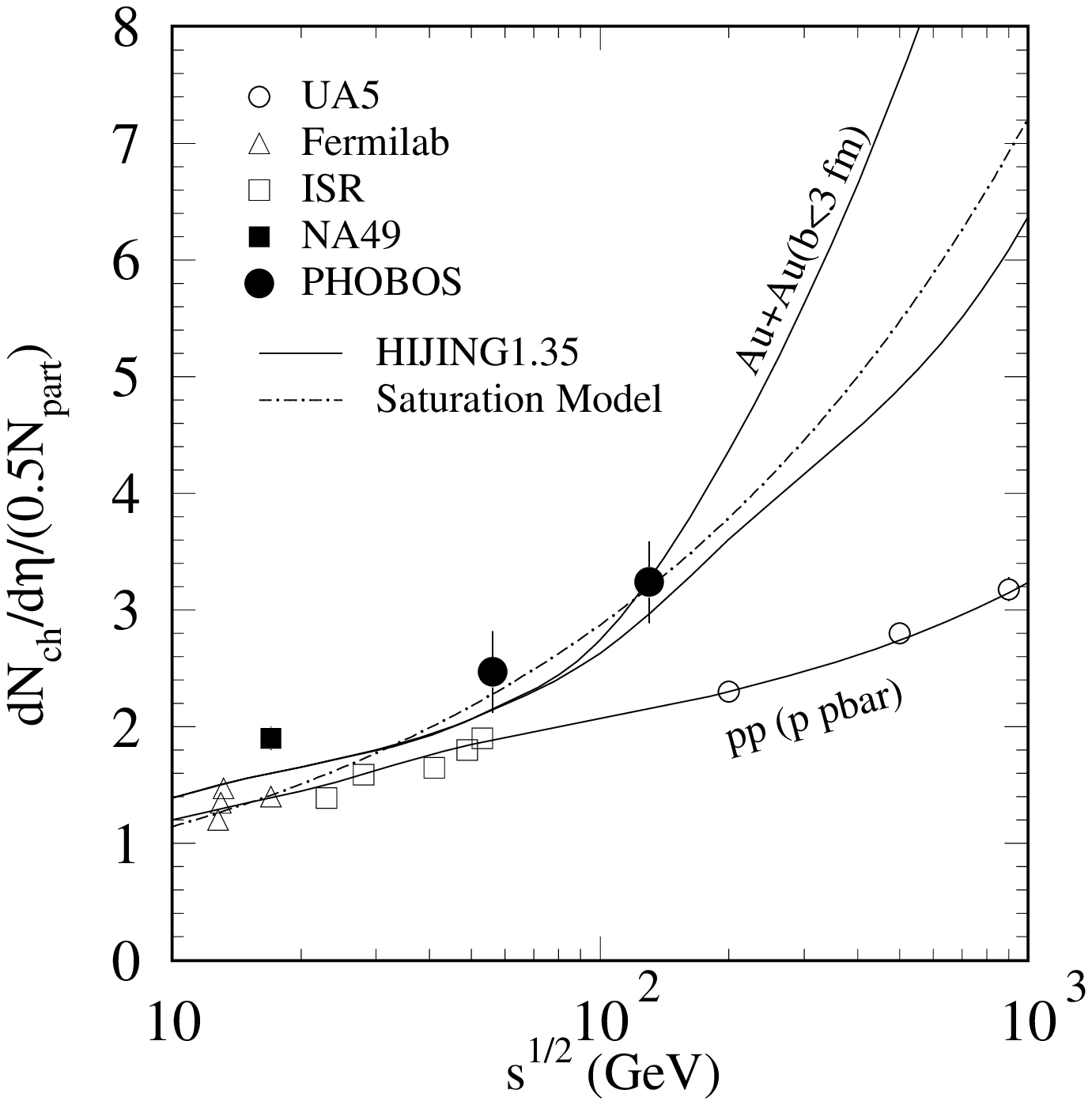}}
\vspace{-7.cm}
\flushright{\epsfxsize=12cm\epsfbox{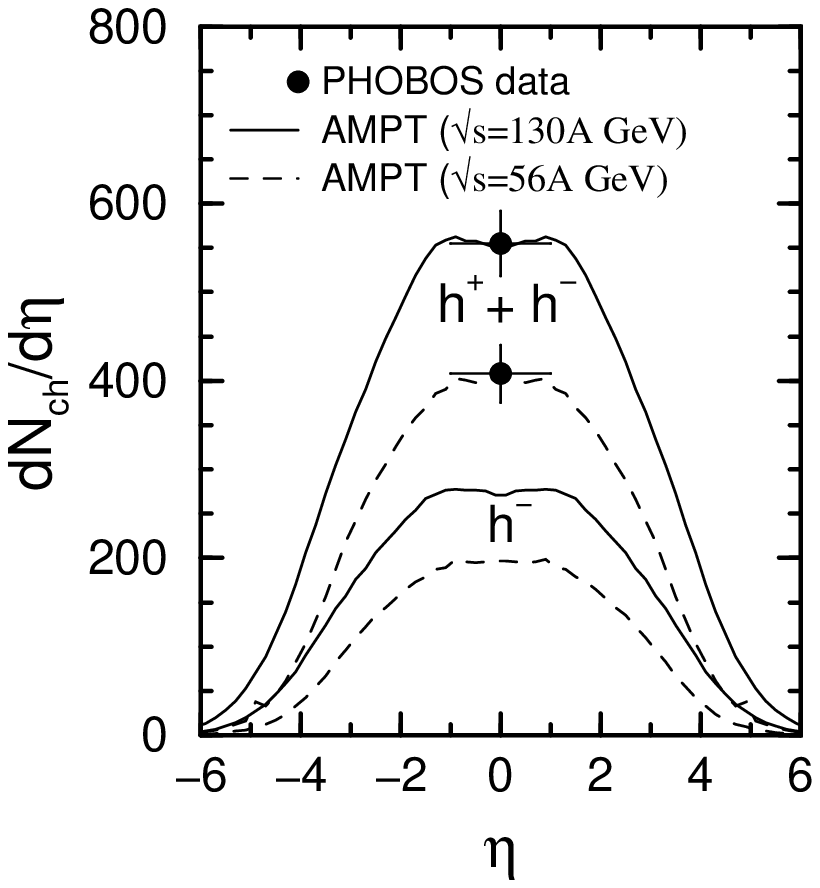}\hspace{-6cm}}
\vspace{-7.8cm}
\caption {\protect\small 
Upper left: LEXUS \cite{LEXUS}
(solid), wounded nucleon model (dashed); 
upper right: Dual String Model with (dashed) 
and without (dotted-dashed) string fusion \cite{DSM+F}; 
lower left: HIJING with shadowing and with and without jet quenching 
(Au+Au, solid) \cite{WG00} together with the EKRT saturation model 
(dotted-dashed) \cite{EKRT}; 
Lower right: AMPT model \cite{AMPT}.}
\vspace{-0.8cm}
\end{figure}

\subsection{Perturbative QCD in $dN_{\rm ch}/d\eta$}

Towards higher cms-energies, perturbative production of gluons and
quarks in $AA$ collisions becomes more important, eventually even dominant, 
as first suggested in \cite{BM87,KLL87,HIJING_91,PCM_91}. The agreement of 
HIJING and EKRT-type models with the first RHIC data suggest that at RHIC 
we are witnessing an onset of pQCD particle production.

In first approximation (leading twist), perturbative production of
partons, minijets, can be computed assuming factorization of the long-
and short-distance physics, i.e.  the nuclear parton distributions
$f_i^A(x,Q)$ and the partonic cross sections $\hat
\sigma_{ij}$. Schematically, the integrated minijet cross section for
producing partons with transverse momenta larger than a cut-off scale
$p_0=1..2$ GeV, is $\sigma_{\rm jet}^{AA}(p_T\ge p_0) = K \sum_{ij} f_i^A
\otimes f_j^A \otimes \hat \sigma_{ij}$.  So far this has been
computed in lowest order pQCD only and a $K$-factor is needed to
simulate the NLO contributions. The good news is, however, that due to
the recent progress \cite{ET00}, the $K$-factor can now be computed
exactly in certain cases, such as in $E_T$ production. The order of
magnitude is $K\sim 2$, but the exact value depends on $\sqrt
s$ and on the parton distributions chosen.

\begin{figure}[hbt]
\vspace{-3.3cm}
\flushleft{ \hspace{-1cm}\epsfxsize=8cm\epsfbox{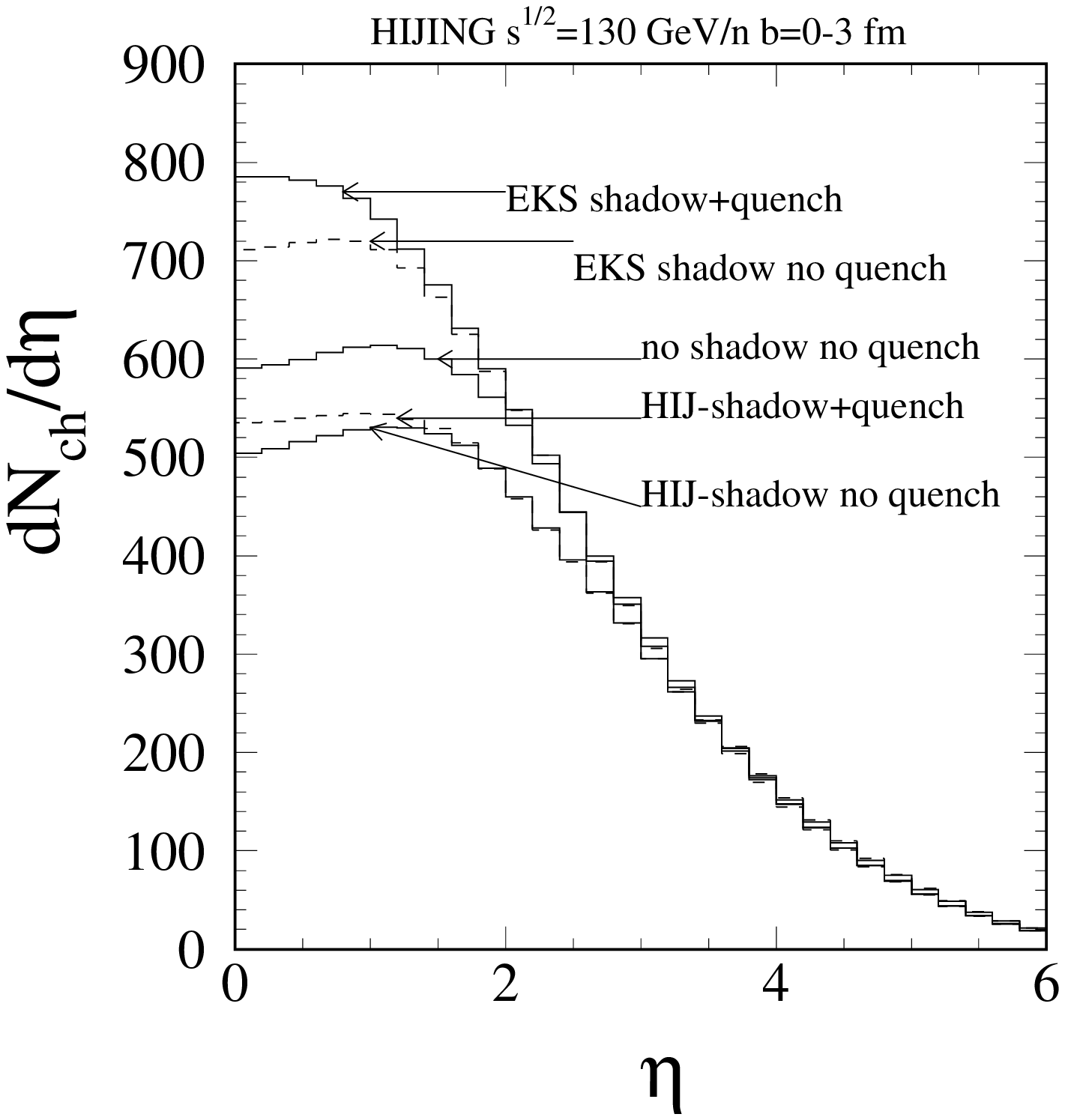}}
\vspace{-0.5cm}
\flushleft{\hspace{0.5cm}\epsfxsize=6cm\epsfbox{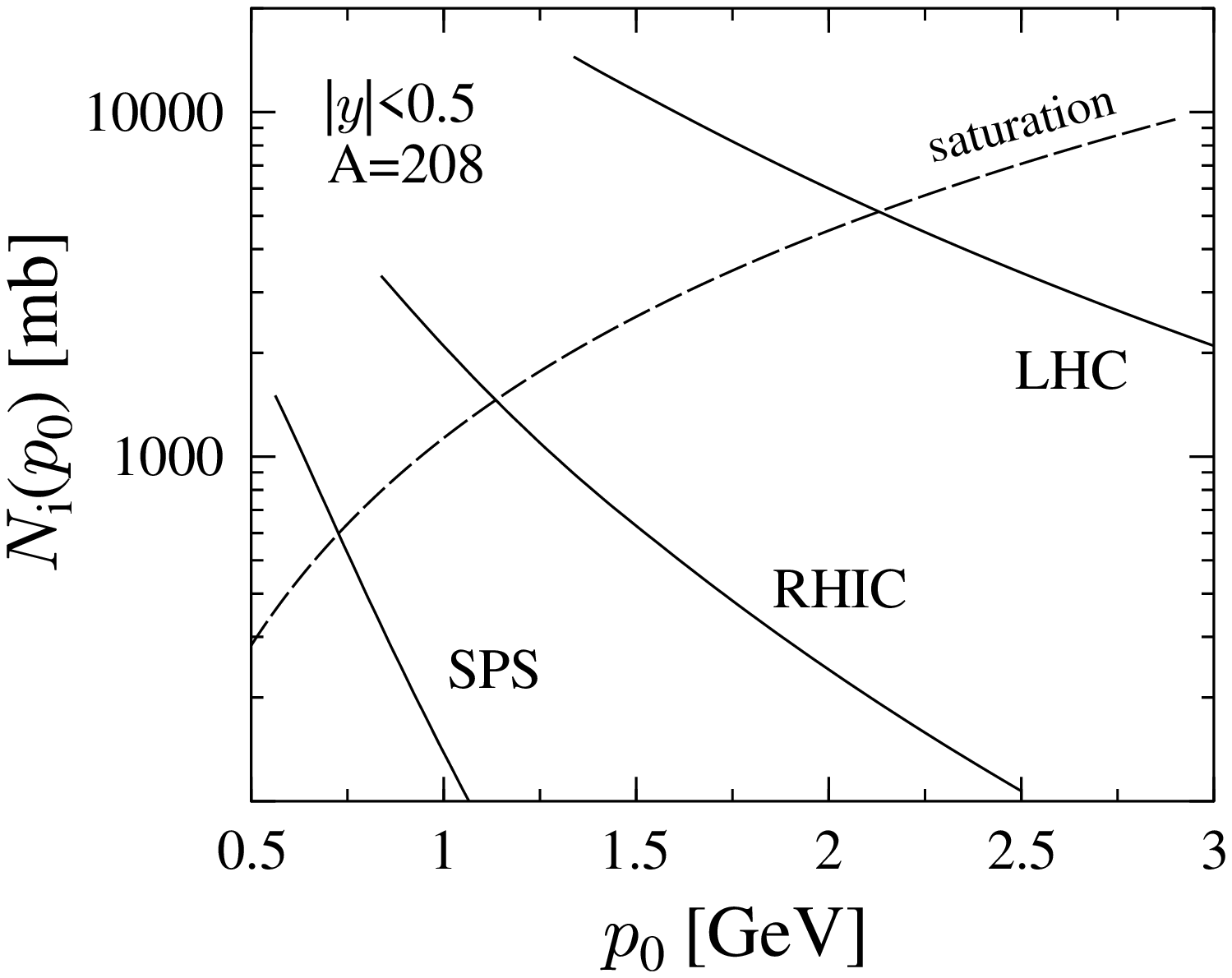}}
\vspace{-16cm}
\flushright{\epsfxsize=8cm\epsfbox{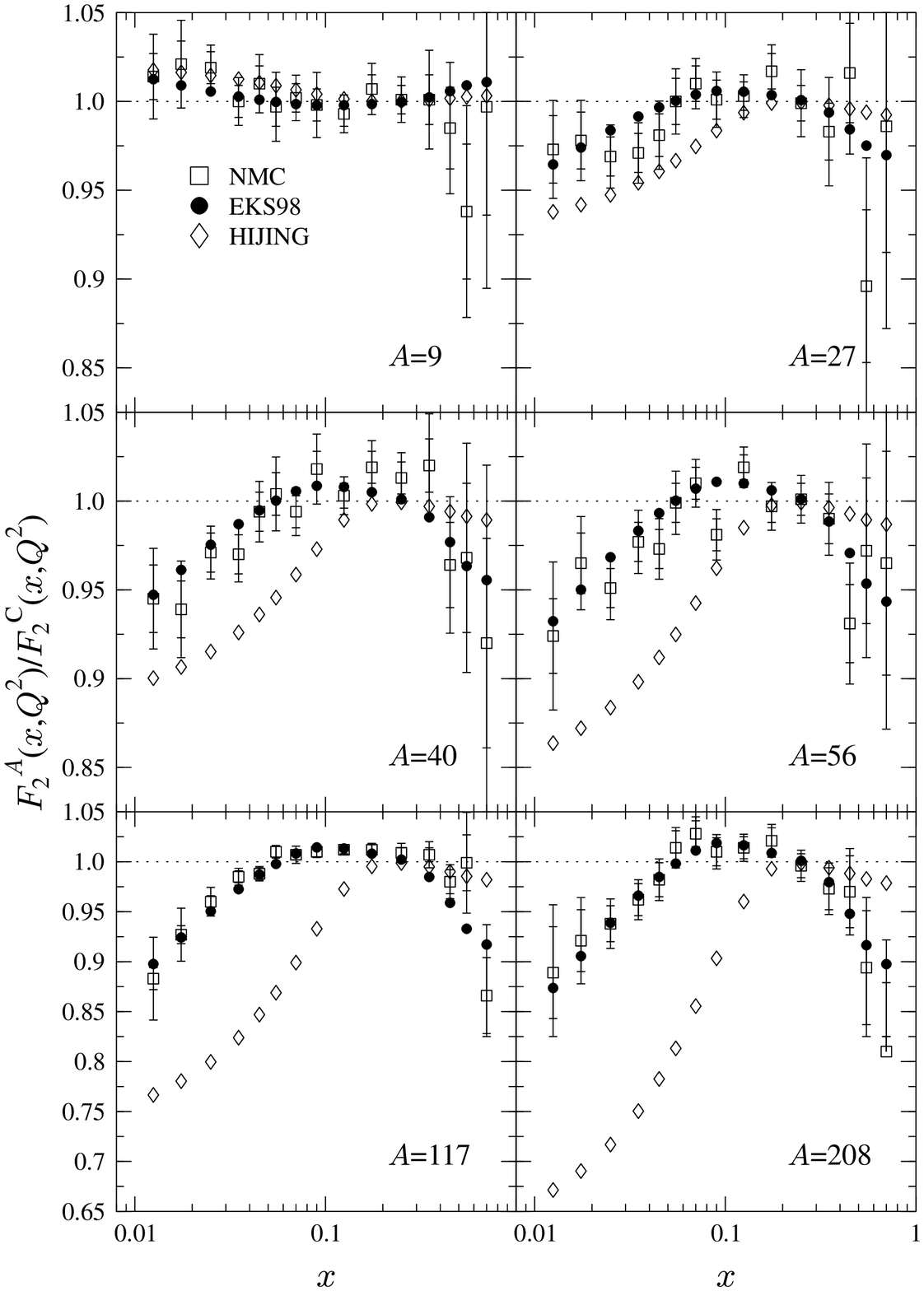}}
\vspace{-0.8cm}
\caption{\protect\small Top left: The HIJING prediction for $dN_{\rm
ch}/d\eta$ in Au-Au collisions at $\sqrt s=130$ $A$GeV
\cite{XNW_private}. Bottom left: Determination of the dynamical
saturation scale $p_{\rm sat}=p_0(A,\sqrt s)$ in the EKRT saturation
model \cite{EKRT}. $N_i$ is the number of produced pQCD partons in
$|y|\le0.5$.  Right: Comparison of the EKS98 nuclear effects (filled
circles) \cite{EKS98} and the HIJING default parameterization (open
diamonds) with the ratio of structure functions $F_2^A/F_2^{\rm Sn}$
measured by NMC \cite{NMC} (open boxes with errorbars).  }
\vspace{-0.7cm}
\label{EKS98_fig} 
\end{figure}

The largest uncertainty in minijet production is the strong dependence
on the minimum transverse momentum scale $p_0$ allowed.  In two-component
models $p_0$ determines the division into the pQCD and
non-perturbative particle production. In e.g. HIJING \cite{HIJING_91},
$p_0=2$ GeV (with $K=2$) is determined from the $pp$ multiplicity
data, and it is kept fixed in the extrapolation to $AA$. This makes
the perturbative component to scale as $N\sim A^{4/3}$.  In
the pQCD models supplemented by the requirement of gluon saturation
\cite{BM87,EKRT}, $p_0$ is determined dynamically as the scale which
dominates all particle production in $AA$ collisions.  This leads to
$p_{\rm sat}=p_0(\sqrt s,A)$, as shown in Fig.~3 (bottom left,
the intersections of the curves) for the EKRT saturation model.

For the modifications of the parton distributions in nuclei,
$f_i^A(x,Q)=R_i^A(x,Q)f_i(x,Q)$, different parametrizations are
available.  The minimum requirement is that a relevant
parametrization is consistent with the existing extensive data
in deeply inelastic $lA$ scattering. If collinear factorization is
used, as usually is the case in models with minijets, the
nuclear modifications should also be compatible with pQCD scale
evolution. The EKS98 nuclear parton distributions \cite{EKS98},
now included also in the CERN-PDFLIB, fulfill these requirements. 
An example of the constraints imposed by the DIS data is shown in 
Fig.~\ref{EKS98_fig}. The effects caused by different 
nuclear parton distributions in the computation of the 
multiplicities in HIJING is shown by Fig.~\ref{EKS98_fig} (top left).

\subsection{Gluon saturation in the initial and final state}

In modeling initial particle production in $AA$ collisions at collider
energies, more emphasis has been recently given to models with gluon
saturation. The idea can be sketched briefly as follows: let
$N_g(Q,\Delta Y)$ gluons appear at a scale $Q$ within a rapidity
correlation length $\Delta Y$.  Saturation of the number of gluons in
the transverse plane takes place when $N_g(Q,\Delta Y)\times
\sigma_g(Q)\sim \pi R_A^2$, where $\sigma_g(Q)\sim \alpha_s(Q^2)/Q^2$
is a cross section for $2\rightarrow 1$ process and $\pi R_A^2$ is the
available total transverse area. Then, provided that $N_g(Q,\Delta Y)$
and $\sigma_g(Q)$ can be computed, the saturation scale $Q_{\rm sat}=Q$ 
which fulfills the saturation criterion, can be found. In the
case of heavy nuclei at high cms-energies, $N_g$ becomes large and
saturation scales become $Q_{\rm sat}\gg\Lambda_{\rm QCD}$.

Gluon saturation {\em in the initial state}, in the wave functions of
the colliding objects, was first discussed in \cite{GLR83} for $pp$
and in \cite{MQ86,BM87} for $AA$ collisions. The idea of initial state
saturation is present also in the classical field approach
\cite{McLV_94,MUELLER99}, where $Q_{\rm
sat}^2=\frac{8\pi^2N_c}{N_c^2-1}\alpha_s xG(x,Q_{\rm sat}^2)T_A(s)$
with the gluon distribution $xG$  and the nuclear overlap
function $T_A(s)$.  The saturation in the final state is obtained through the
initial state gluon saturation, as the transverse profile of the
rapidity density of produced gluons in a central $AA$ collision can be
expressed as \cite{MUELLER99} ${dN_g^{AA}}/{d^2sdy}=
c{C_FQ_{\rm sat}^2}/(\alpha_s2\pi^2)$. The ``gluon liberation
constant'' $c$ is expected to be of the order of one \cite{MUELLER99}.
Analytical calculations in the semiclassical approximation 
\cite{KOVCHEGOV00} give $c=1.39$, and it is interesting to note that if an agreement with the PHOBOS data is
required, one gets $Q^2_{\rm sat}=2.1$ GeV$^2\gg\Lambda_{\rm QCD}^2$. The
SU(2) lattice formulation of the classical field approach \cite{KV00}
gives $c=1.29\pm0.09$, the SU(3) calculations are in progress.
Similar values are also obtained in \cite{KN00}, based on fits to the
PHOBOS data.

Saturation {\em in the final state}, i.e. of produced gluons is also
possible, even without the requirement of saturation in the initial
state. In the EKRT saturation model \cite{EKRT} the number of produced
gluons above a minimum $p_T$-scale $p_0$, $N_{AA}(p_0,\sqrt s,A)$, is
computed from pQCD. The cross section in the saturation criterion is
simply taken as the geometrical one, $\sigma_g=\pi/p_0^2$, not
specifying any powers of $\alpha_s$ or group theoretical
factors.\footnote{The suppression of the production of small-$p_T$
quanta, leading to the saturation, can also be viewed in terms of a
screening mass generated into the produced system \cite{EMW}.} The
saturation criterion in a central rapidity unit of central collisions,
$N_{AA}(p_0,\sqrt s)\times\pi/p_0^2=\pi R_A^2$, then determines the
dominant scale $p_{\rm sat}=p_0(\sqrt s,A)$ as shown in
Fig.~\ref{EKS98_fig}.  At a scaling limit $\sigma_{\rm jet}\sim
p_0^{-2}$, the saturation scale behaves as $p_{\rm sat}^2\sim A^{1/3}$,
causing the initial state multiplicity to grow as $N_{AA}\sim A$
instead of $A^{4/3}$. A full calculation with realistic parton
distributions accounting for the small-$x$ increase and the EKS98
nuclear effects \cite{EKS98} gives a multiplicity of produced partons
as \cite{EKRT} $N_{AA}(p_{\rm sat})= 1.383A^{0.922}(\sqrt s)^{0.383}$.

Also the released transverse energy, $E_{Ti}^{AA}(p_{\rm sat})$ at
saturation can be computed from pQCD. As the formation time of
the produced system is  $\tau_i\sim 1/p_{\rm sat}$, the
initial densities in the EKRT approach can be computed through
converting $E_{Ti}^{AA}(p_{\rm sat})$ into a thermal energy
density $\epsilon_i$, which further converts into a thermal number
density $n_i^{\rm th}$.  On the other hand, conversion of the initial
number of gluons at saturation, $N_{AA}(p_{\rm sat})$, into $n_i$ gives
densities very close to the thermal $n_i^{\rm th}$.  In this sense the
system looks thermal already at saturation \cite{EKRT}. Thus assuming 
thermalization at $\tau_i$, and an isentropic expansion stage
described in terms of (boost invariant) hydrodynamics, the final state
multiplicity of charged particles is computable from the entropy
produced initially at $\tau_i$.  This leads to a simple scaling law
\cite{EKRT} $dN_{\rm
ch}/dy\approx\frac{2}{3}\times\frac{3.6}{4}N_i=0.83A^{0.922}(\sqrt
s)^{0.383}$ for central collisions.  Agreement with the PHOBOS data is
amazingly good, as seen in Fig.~2 (lower left).  More detailed
description of the expansion stage in terms of hydrodynamics with
transverse flow effects and resonance decays has now also been done 
\cite{ERRT}, the results are shown in Fig.~4.

\begin{figure}[hbt]
\vspace{-0.6cm}
\begin{minipage}{6.5cm}
{\protect\small Figure 4. $dN_{\rm ch}/d\eta$ and $dE_T/d\eta$
averaged in $|\eta|\le1$ vs. $\sqrt s$, as obtained from
pQCD+saturation+2d hydrodynamics+resonance decays \cite{ERRT}. The
open symbols are for central collisions, the closed ones are with a 6\%
centrality selection. In the left panel, the lines are from \cite{EKRT} 
for central collisions, the open circles are the PHOBOS
data \cite{PHOBOS}.  }
\end{minipage}
\vspace{-6.0cm}
\flushright{
\epsfxsize=6cm\epsfbox{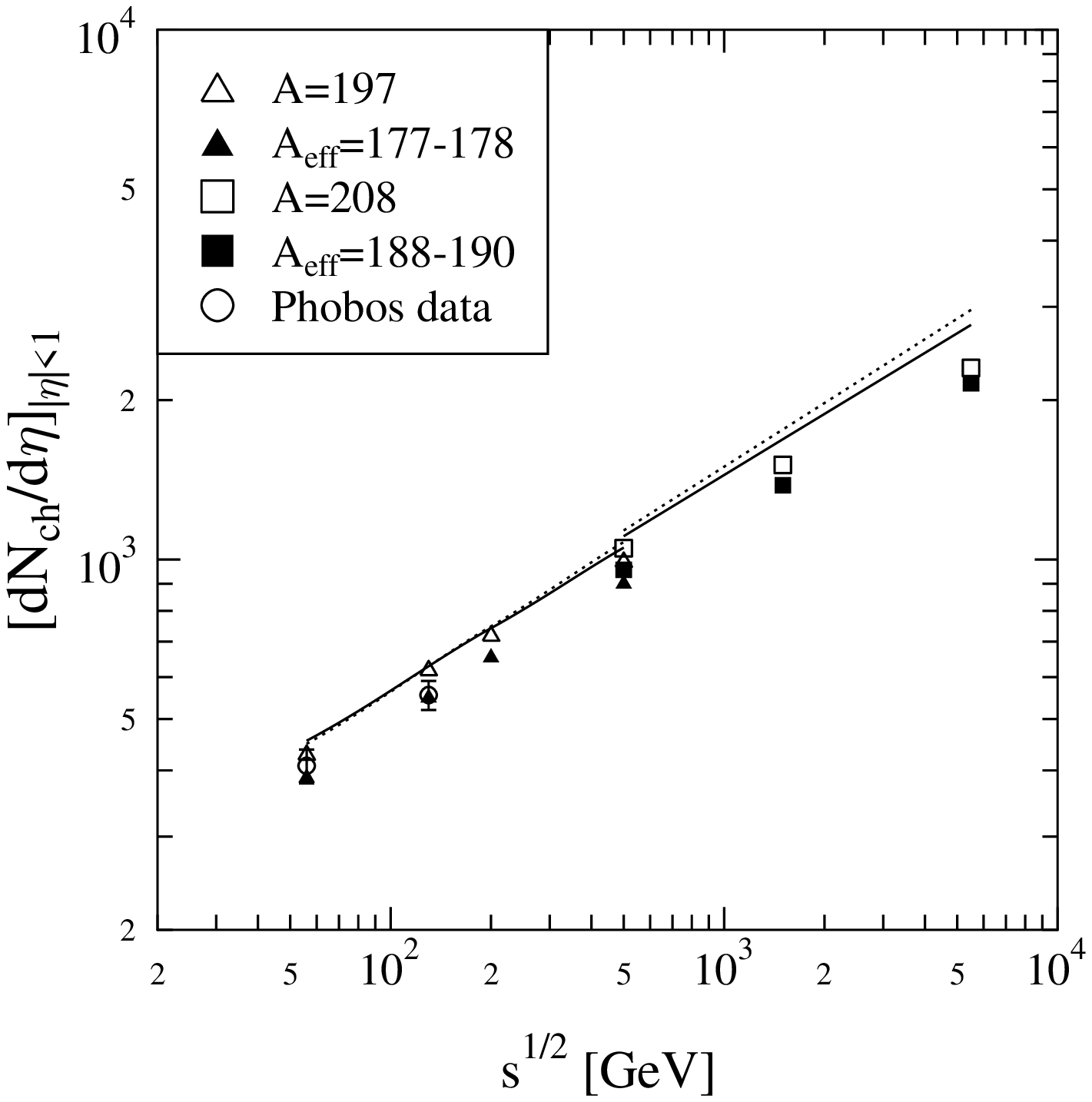}\hspace{-1.2cm}
\epsfxsize=6cm\epsfbox{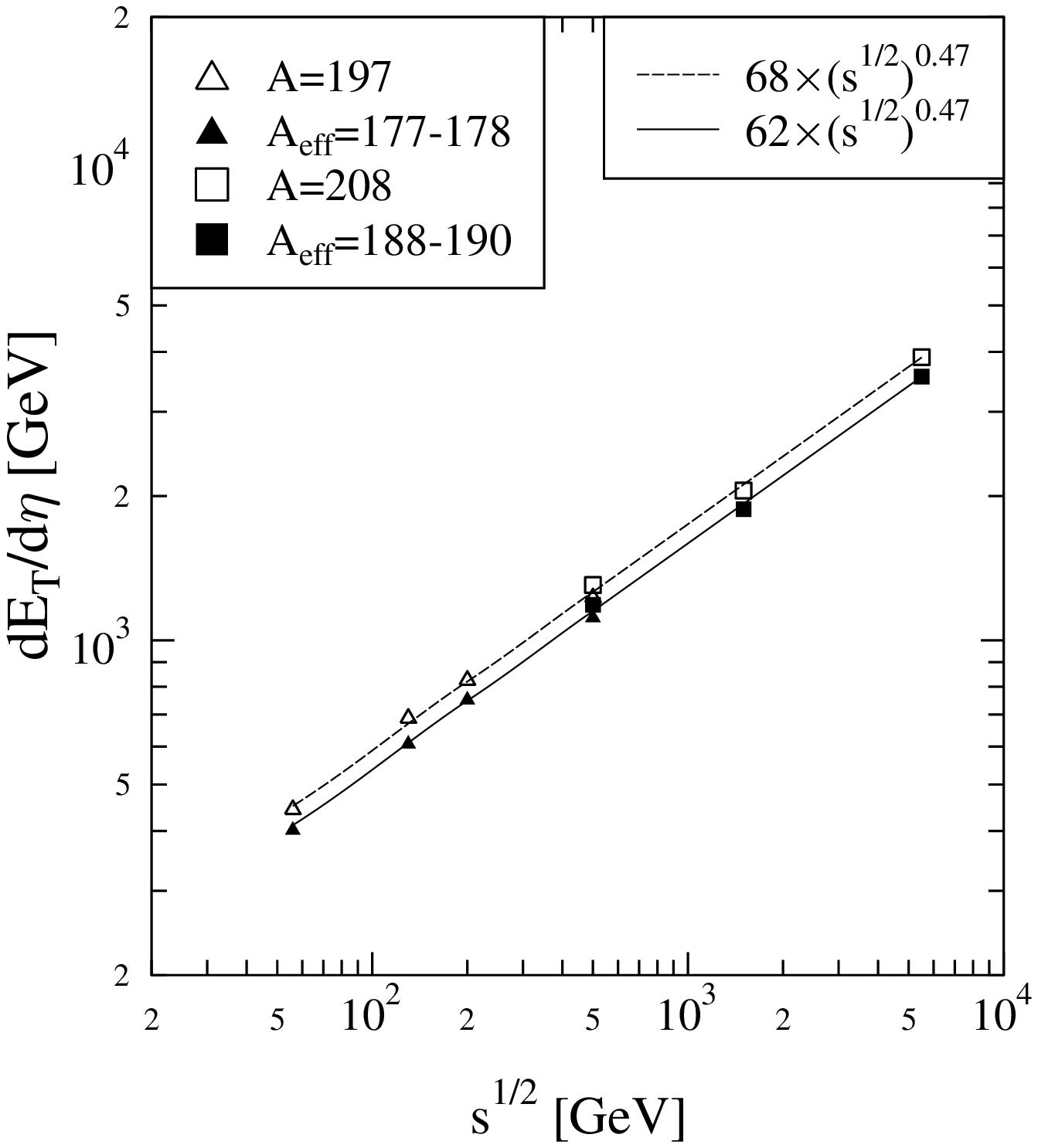}\hspace{-2cm}
}
\label{RVESA}
\vspace{-1.5cm}
\end{figure}

\section{Centrality dependence of $dN_{\rm ch}/d\eta$}

The dependence of $dN_{\rm ch}/d\eta$ on the centrality of the
collision has been suggested in \cite{WG00} for testing the predictions
from different models. Before the release of the PHENIX data on the 
centrality dependence, three concrete predictions were submitted: 

1. In HIJING \cite{WG00} the multiplicity is formed as a combination
of a component which scales as the number of participants $\sim A$, and a pQCD
component which scales with the number of binary collisions $\sim A^{4/3}$ 
(keeping $p_0$ fixed): $dN_{\rm ch}({\bf b})/d\eta = N_{\rm part}({\bf
b})n_{\rm soft} + f N_{\rm bin}({\bf b})\sigma_{\rm jet}^{AA}(p_0)$

2. The EKRT saturation model \cite{EKRT} can be extended to describe
the saturation of gluons locally at a transverse distance {\bf s} in a
collision with an impact parameter {\bf b}. Assuming the extreme case
that saturation fully dominates the particle production, one obtains
\cite{EKT00} $dN_{\rm ch}({\bf b})/d\eta\approx 0.9\times0.9\times \frac{2}{3}
\times \int d^2{\bf s} \, p_{\rm sat}^2(\sqrt s, A, {\bf b}, {\bf
s})/\pi$.

3. Dividing the particle production into a ``hard'' and a ``soft''
components which scale as $\sim N_{\rm part}$ and $\sim N_{\rm bin}$,
correspondingly, one may write \cite{KN00} $dN_{\rm ch}({\bf b})/d\eta
= xn_{pp}N_{\rm part}({\bf b}) + (1-x)n_{pp}N_{\rm bin}({\bf b})/2$.
The average particle multiplicity in $n_{pp}$ is obtained from the $pp$
data, and the fraction $x\sim0.05\dots0.09$ is obtained by fits to the
PHOBOS data \cite{PHOBOS}. In \cite{KN00}, also an extension of the
initial state saturation model \cite{MUELLER99} to noncentral
collisions was suggested in the form $dN_{\rm ch}/d\eta = \frac{2}{3}c
N_{\rm part} xG(x,Q_{\rm sat}({\bf b})^2)$, with $c$ obtained from a fit 
to PHOBOS data.

\begin{figure}[hbt]
\vspace{-0.5cm}
\centerline{\epsfxsize=5cm\epsfbox{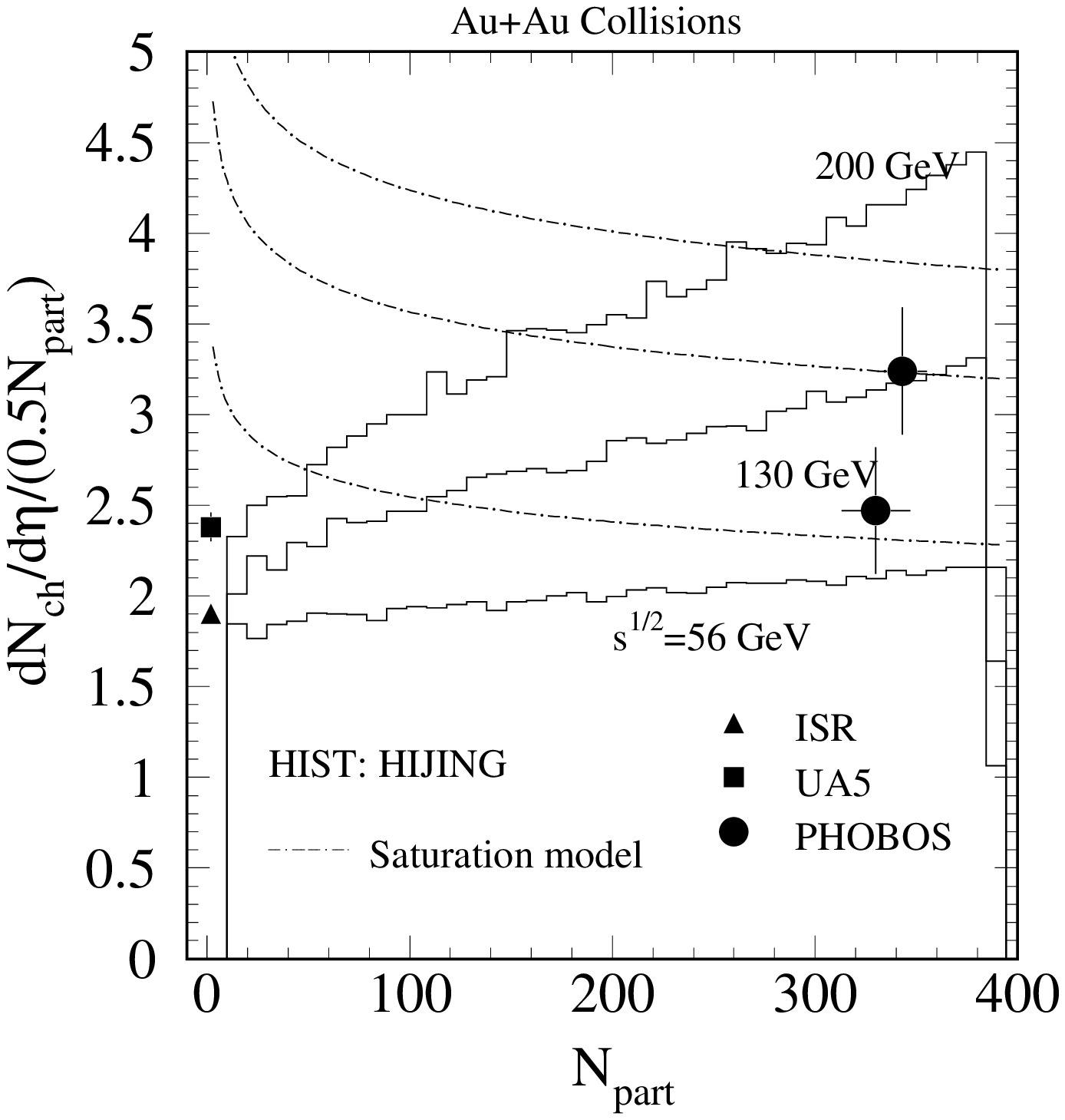} \hspace{12cm}}
\vspace{-6cm}
\centerline{\epsfxsize=7cm\epsfbox{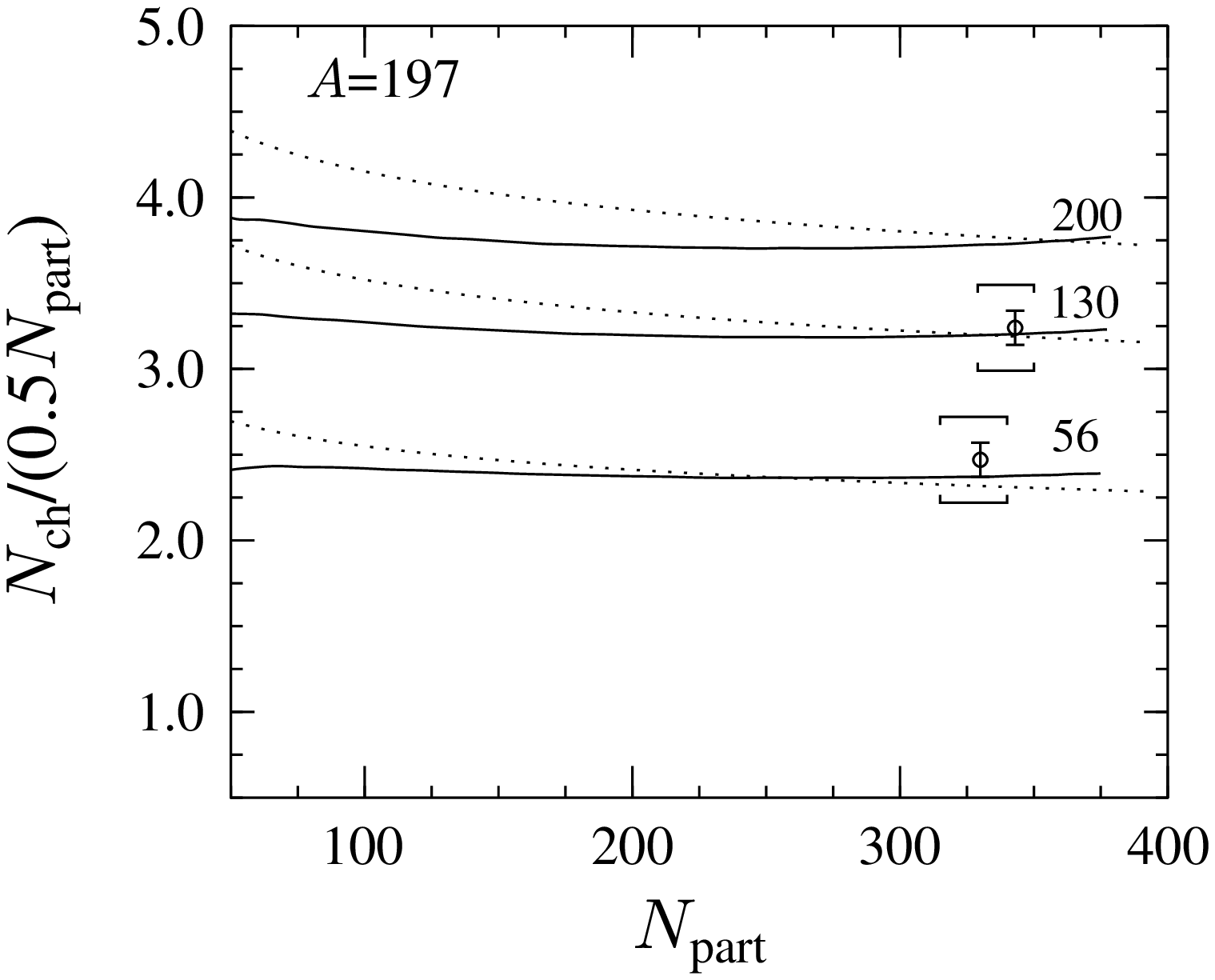}\hspace{1cm}}
\vspace{-7cm}
\centerline{\epsfxsize=4.5cm\epsfbox{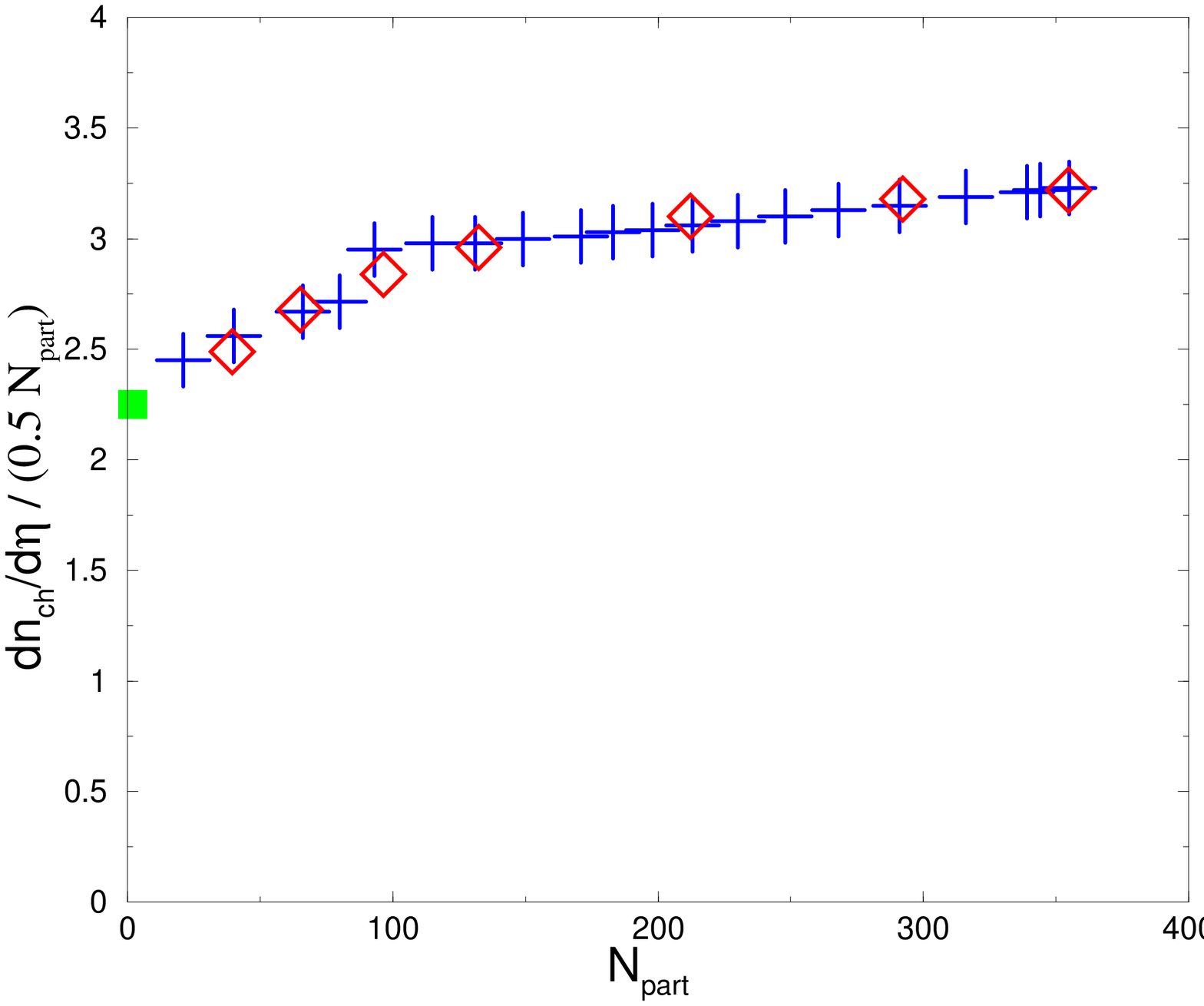}\hspace{-10.5cm}}
\vspace{0cm}
\centerline{\epsfxsize=5.5cm\epsfbox{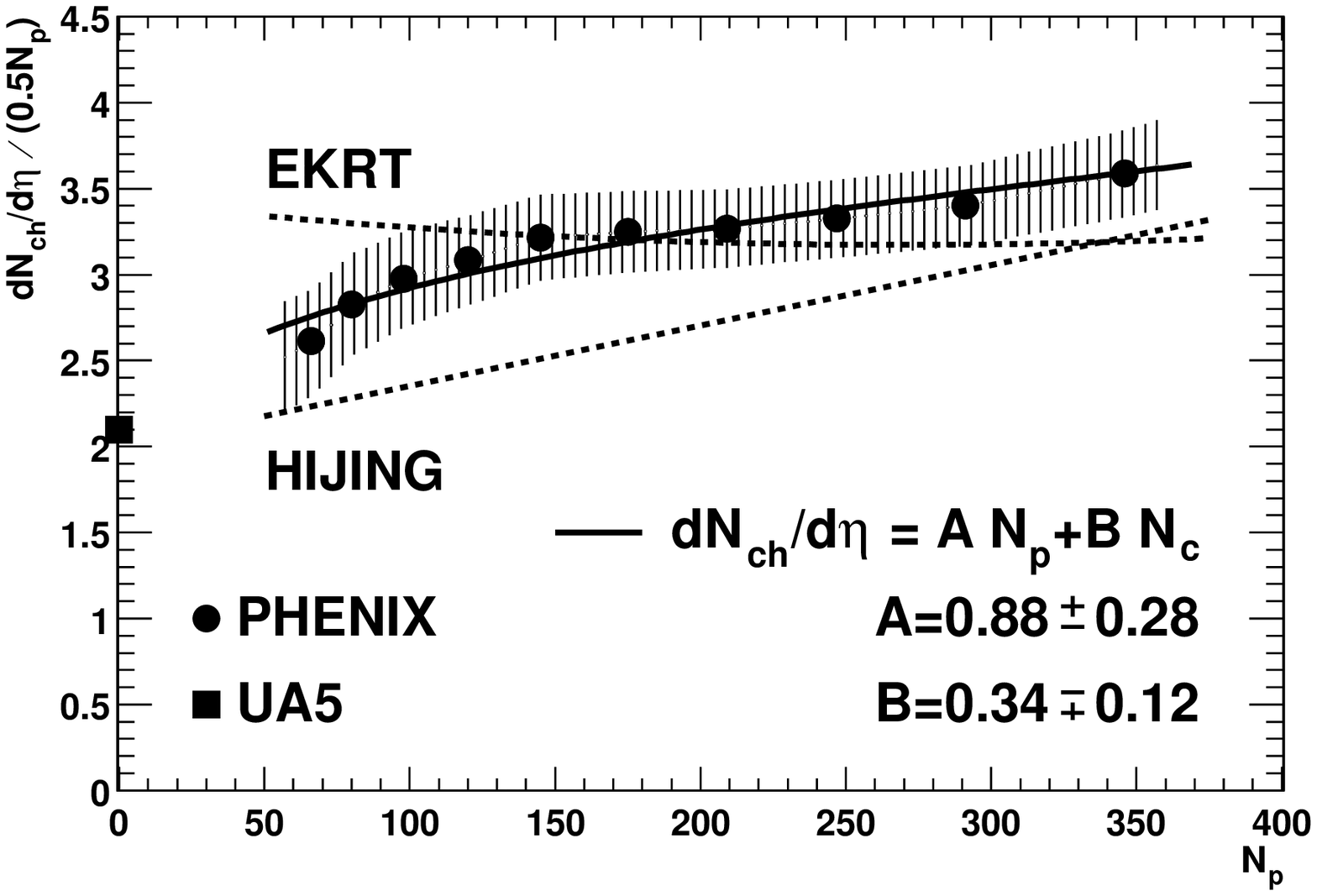}\hspace{-11.5cm}}
\vspace{-3cm}
\begin{minipage}{10cm}
Figure 5. {\small Multiplicity $dN_{\rm ch}/d\eta/(0.5N_{\rm part})$
vs. $N_{\rm part}$ at $\sqrt s=130$ $A$GeV. From left: HIJING
\cite{WG00} (histograms); local saturation model \cite{EKT00} (solid)
and EKRT saturation \cite{EKRT} (dotted); ``hard''+''soft'' eikonal
approach (crosses) and initial state saturation model (diamonds) from
\cite{KN00}.  Lower right: the PHENIX data \cite{PHENIX}.  }
\end{minipage}
\label{centrality_fig}
\vspace{-0.8cm}
\end{figure}

As shown in Fig.~5 (lower right), the centrality data measured by
PHENIX \cite{PHENIX} lie in between the EKRT type saturation model and
HIJING, quite close to the behaviour obtained in \cite{KN00}.
Obviously, gluon saturation (EKRT) is not the dominant mechanism for
non-central ($N_{\rm part}\lsim200$) collisions at $\sqrt s=130$
$A$GeV but a non-saturated component is needed at small transverse
densities.  For central collisions, however, gluon saturation is not
ruled out as the dominant mechanism, as also indicated by the PHOBOS data.

The multiplicity measurements alone are not sufficient to rule out
models, as most models can be tuned to fit the measured multiplicities
and even their centrality dependence. For instance, HIJING has
theoretical uncertainties in the division into soft and hard physics
and in the extrapolation to $AA$, and also in the pQCD component as
discussed above. The saturation models in turn suffer from uncertainties
related to the saturation criterion which may be of non-perturbative 
nature and contain unknown constants and powers of (running) $\alpha_s$ 
which may need to be phenomenologically determined. In any case, it is
unlikely that the true theoretical errors in any of the models 
could be squeezed below the $\pm10$\% level already reached by the first 
RHIC data.

Of the global variables, $E_T$ should be more efficient in ruling out
models than $N_{\rm ch}$, as $E_T$ is much more sensitive to the
evolution of the system: as shown in \cite{EKRT,ERRT}, the ratio of
the initially released $E_T$ (at saturation) to the measurable final
state $E_T$ is roughly a factor three. The reduction is due to the $pdV$
work during the expansion stage. Such reduction does not take place in
models which do not describe the final state interactions
(e.g. HIJING, apart form jet quenching). Also interestingly, the
amount of $E_T$ released in the classical field model \cite{KV00} is
twice the initial $E_T$ obtained in \cite{EKRT}, so obviously a very
strong energy loss  would be needed in order to fall near the 
measurements of $E_T$ shown by PHENIX in this conference.

\section{Elliptic flow}

In non-central collisions, the anisotropic flow has been suggested as
a signature of formation of pressure and as a probe of the early
stages of the produced system \cite{OLLITRAULT92}.  Pressure can only
be formed if the produced particles interact with each other.  In
non-central collisions the spatial azimuthal asymmetry of the
production zone is transmitted to the pressure gradients which drive
the transverse flow.  In non-central collisions, the initially
generated flow is asymmetric, $\langle v_x^2\rangle \ne
\langle v_y^2\rangle$, leading eventually to an asymmetry in the
azimuthal particle distributions. The harmonic coefficients of the
Fourier expansion of the particle spectra are a quantitative measure
of this asymmetry.  In particular, the elliptic flow is
characterized by the coefficient $v_2$, defined as
\begin{equation}
v_2(y) = \frac{\int d\phi \cos(2\phi)\frac{dN}{dyd\phi}} { \int d\phi
\frac{dN}{dyd\phi}} \quad\quad {\rm and}
\quad\quad\quad 
v_2(y,p_T) = \frac{\int d\phi
\cos(2\phi)\frac{dN}{dydp_T^2d\phi}} { \int d\phi
\frac{dN}{dydp_T^2d\phi}}.
\end{equation}

Clear elliptic flow signal, up to 6\% for $v_2$ and up to 15 \% for
$v_2(p_T)$ in minimum bias events in Au-Au at $\sqrt s=130$ $A$GeV,
has now been measured at RHIC by STAR \cite{v2_STAR} (shown in
Fig.~\ref{v2_hydro}). An attempt to briefly summarize the situation
with the semiclassical cascade models is presented in Fig.~6, where
$v_2$ is shown from RQMDv2.4, UrQMD and ZPC at $\sqrt s=200$ $A$GeV
(the measured $v_2$ can be expected to be close to that at $\sqrt
s=200$ $A$GeV \cite{KSH00}): Partonic cascades can in principle
generate large enough $v_2$ but very large rescattering cross sections
are needed \cite{v2_ZPC99}. As the mean-free paths of particles then
become very short, this speaks in favour of hydrodynamics. Hadronic
cascades predict the linear rise of $v_2(p_T)$ at $p_T\lsim500$ MeV
but they level off too soon after that \cite{v2_UrQMD00}.  These
models do not seem to produce sufficient $p_T$-averaged $v_2$ either
\cite{v2_UrQMD00,v2_RQMDv2.4}, which speaks in favour of an early
formation of collectivity and pressure in the partonic system.

\begin{figure}[hbt]
\vspace{-1.5cm}
\centerline{\hspace{0cm}
\epsfxsize=5cm\epsfbox{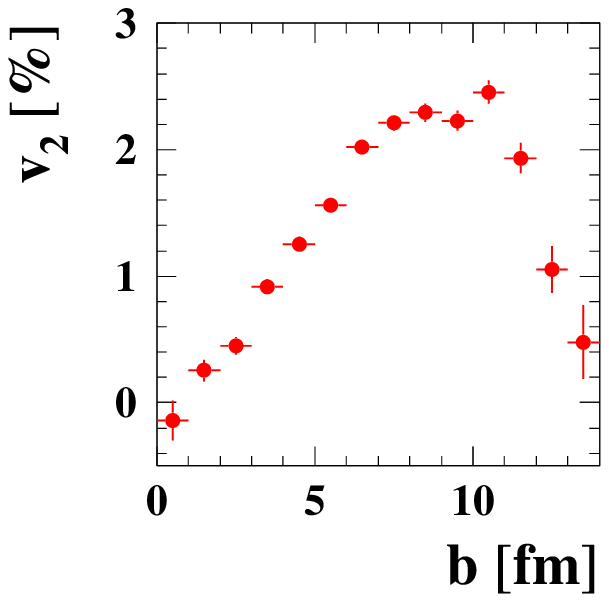}\hspace{0.5cm}
\epsfxsize=6cm\epsfbox{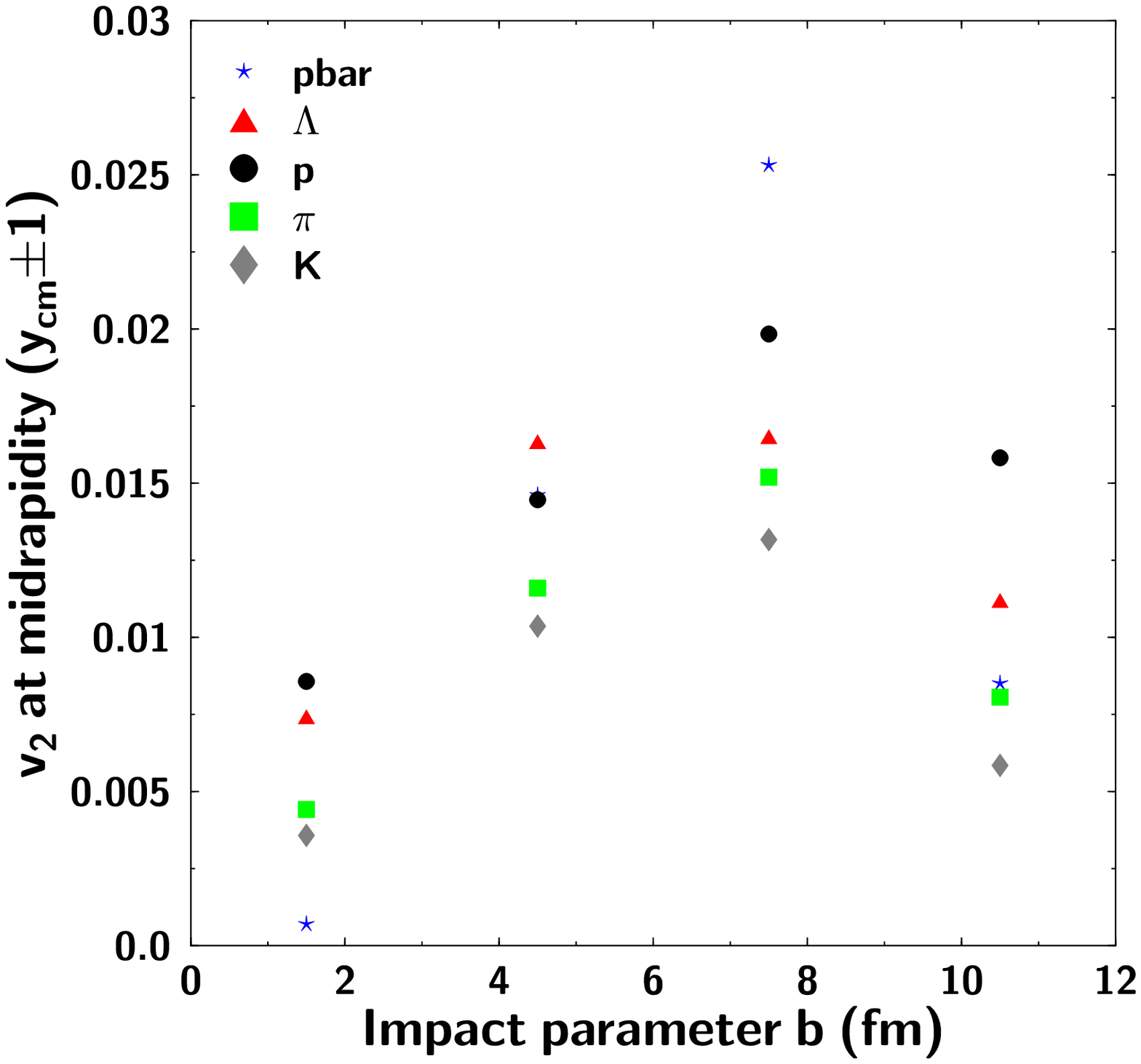}\hspace{-0.5cm}
\epsfxsize=6cm\epsfbox{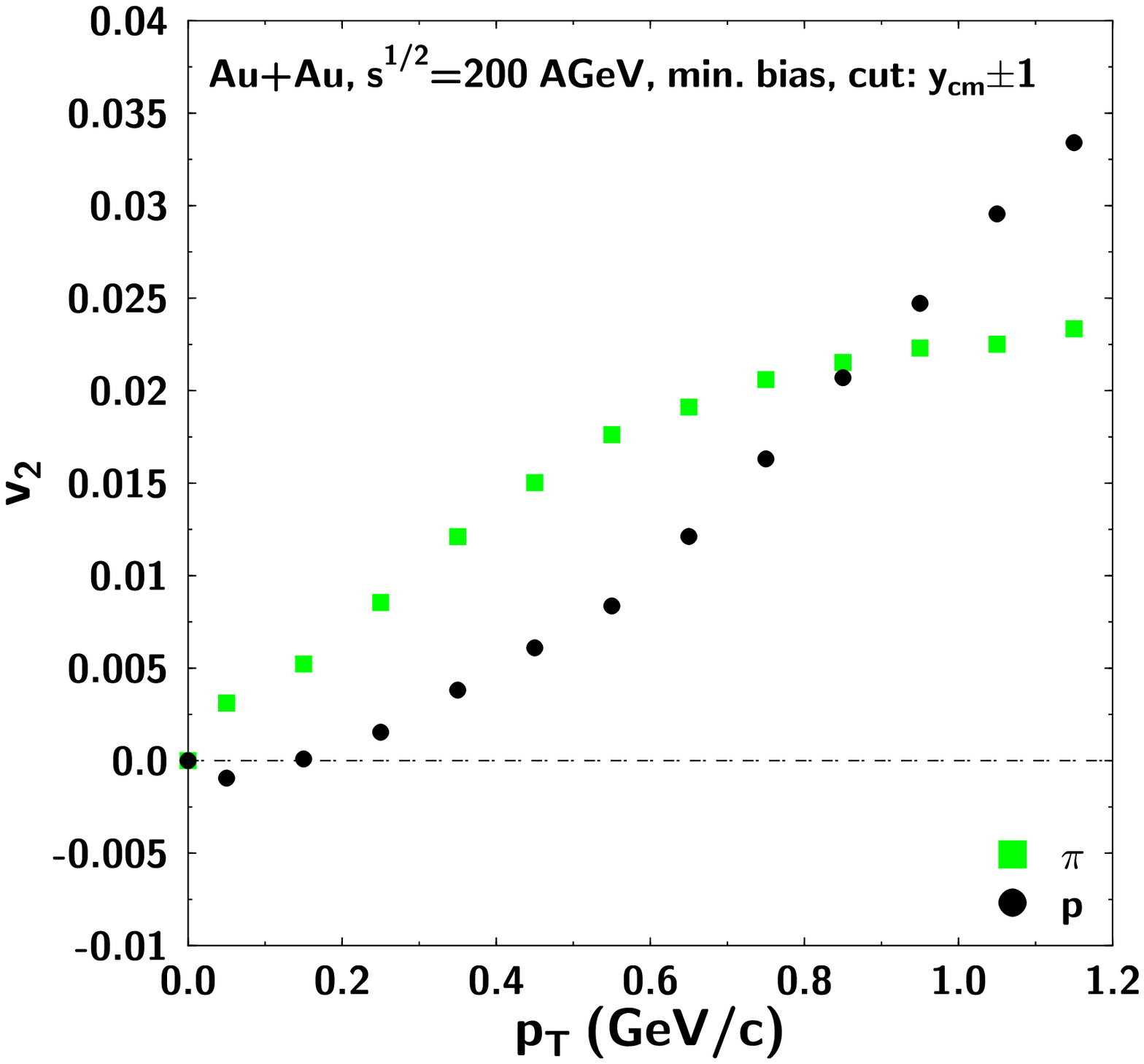}\hspace{0cm}
}
\centerline{ \hspace{-11cm}
\epsfxsize=4cm\epsfbox{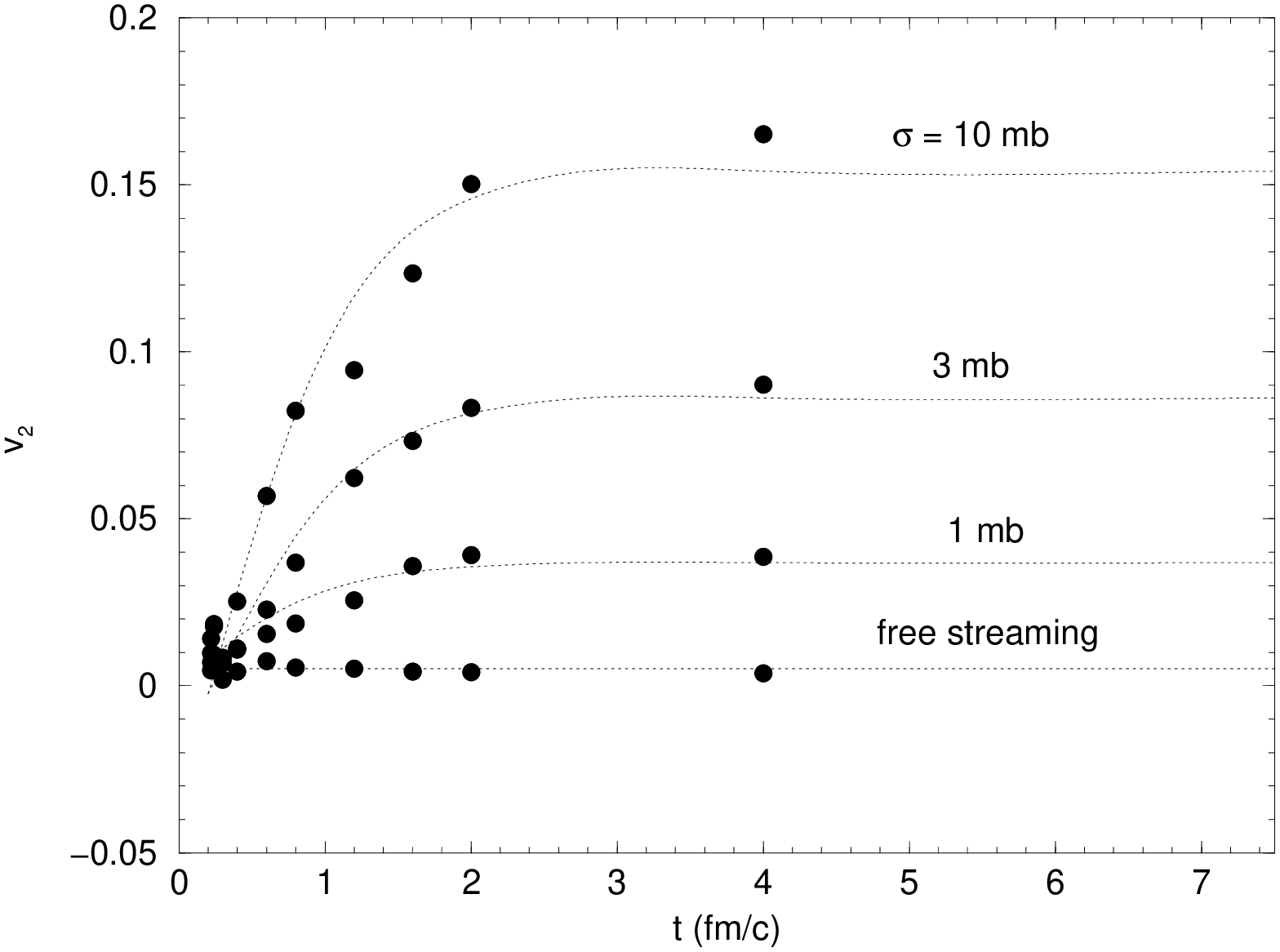}\hspace{0cm}}
\vspace{-2.5cm}\hspace{5.5cm}
\begin{minipage}{10cm}
Figure 6. {\small Elliptic flow for Au-Au at $\sqrt s=200$ $A$GeV from
cascade models.  From upper left: $v_2$ vs. {\bf b} from RQMDv2.4
\cite{v2_RQMDv2.4}, $v_2$ vs. {\bf b} from UrQMD \cite{v2_UrQMD00},
and $v_2(p_T)$ vs. $p_T$ from UrQMD vs. $p_T$ \cite{v2_UrQMD00}.
Lower panel: $v_2$ vs. time from the ZPC parton cascade
\cite{v2_ZPC99}.  }
\end{minipage}
\label{v2_cascades}
\vspace{-0.5cm}
\end{figure}

The observed strong elliptic flow confirms the prediction based on
hydrodynamics \cite{KSH00}. Detailed comparisons have also been done,
as shown in Fig.~\ref{v2_hydro}.  The $v_2$ obtained in the
``conventional'' hydrodynamic approach \cite{KHHH00}, where the
initial conditions are constrained by the observed final state
multiplicity and where the Cooper-Frye procedure \cite{CF} is applied
for the decoupling, agrees with the data quite well.  Similar
conclusion is obtained in the Hydro-to-Hadrons approach \cite{TLS00},
where the RQMDv2.4 hadron cascade is switched on below $T=160$ MeV in
order to treat the chemical and kinetic freeze-out in more detail.  At
very large impact parameters, where application of a purely
hydrodynamical system may be questioned in any case, these approaches
seem to produce too much $v_2$. Also $v_2(p_T)$ in the conventional
hydrodynamic approach agrees with the STAR data very well up to 2 GeV
or so. The deviation at high $p_T$ is not surprising, since in a
finite dynamic system with a finite lifetime the large-$p_T$ tails
will not be adequately described by hydrodynamics. Another interesting
observation is that the amount of $v_2$ is not very sensitive to the
equation of state (EoS) used. Therefore, the single particle spectra are
needed to constrain the remaining uncertainties such as the EoS.

\setcounter{figure}{6}
\begin{figure}[hbt]
\vspace{-0.5cm}
\flushleft{
\epsfxsize=5cm\epsfbox{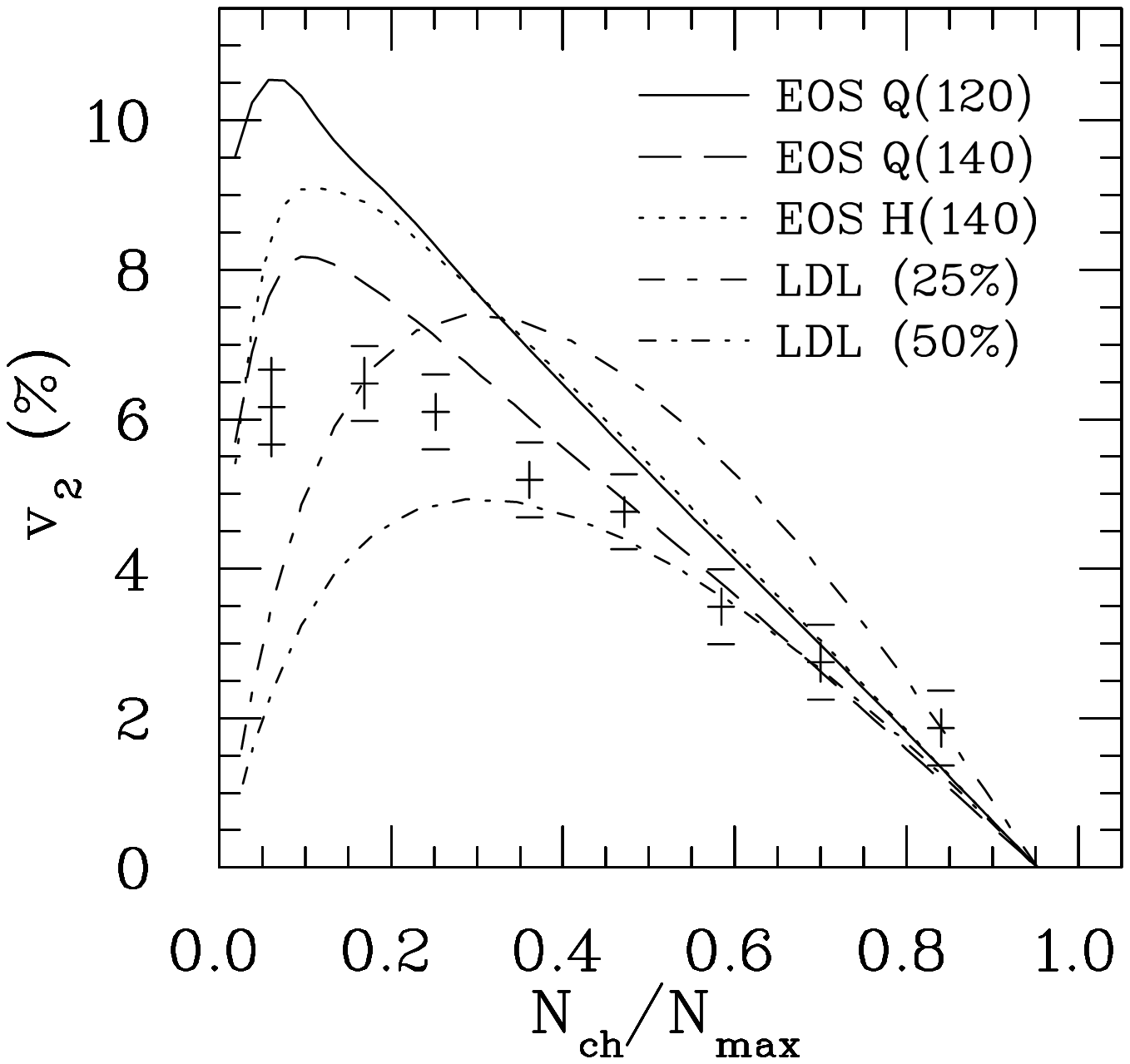}\hspace{0.2cm}
\epsfxsize=5cm\epsfbox{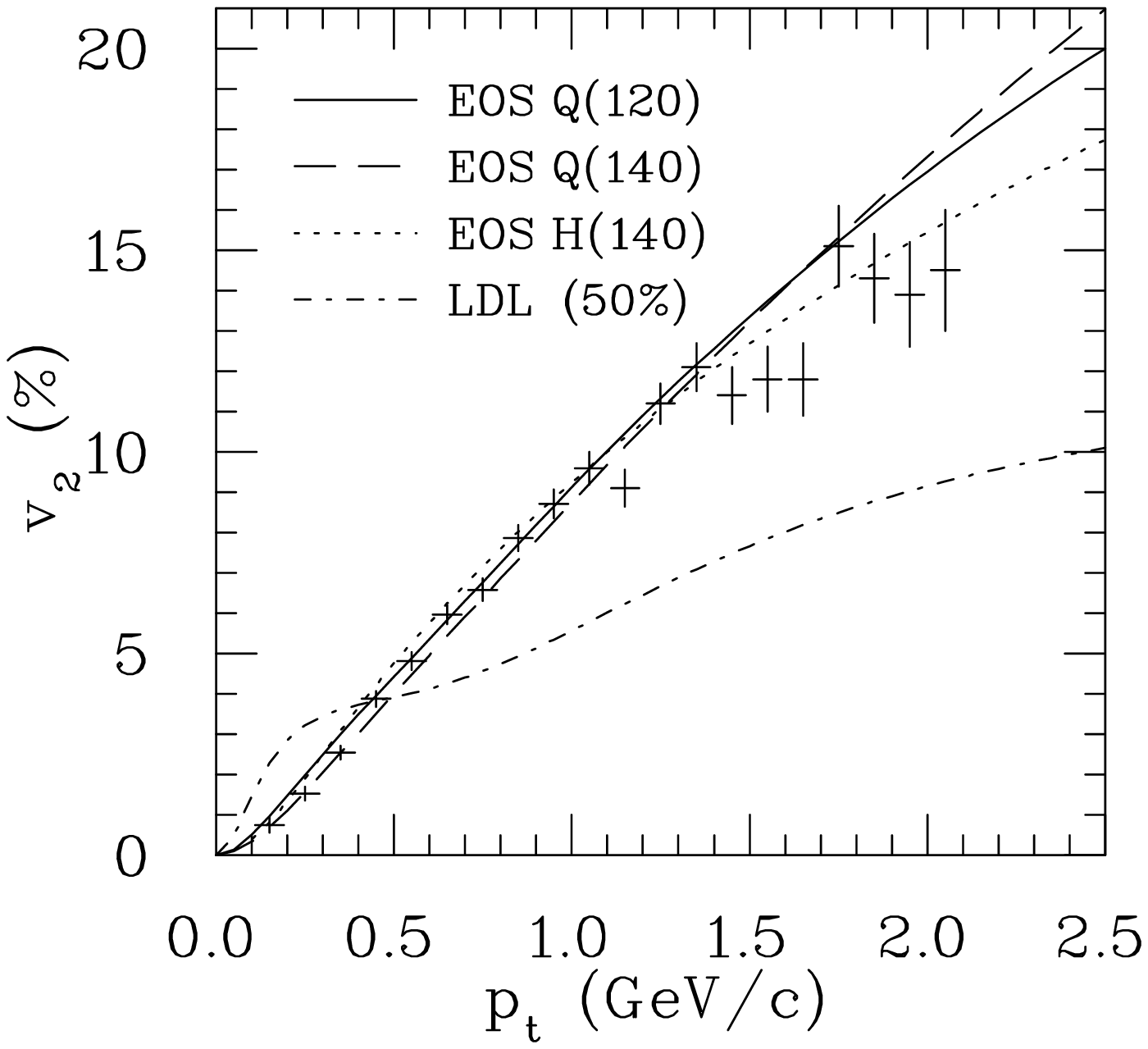} }
\vspace{-5.7cm}
\flushright{\epsfxsize=5.5cm\epsfbox{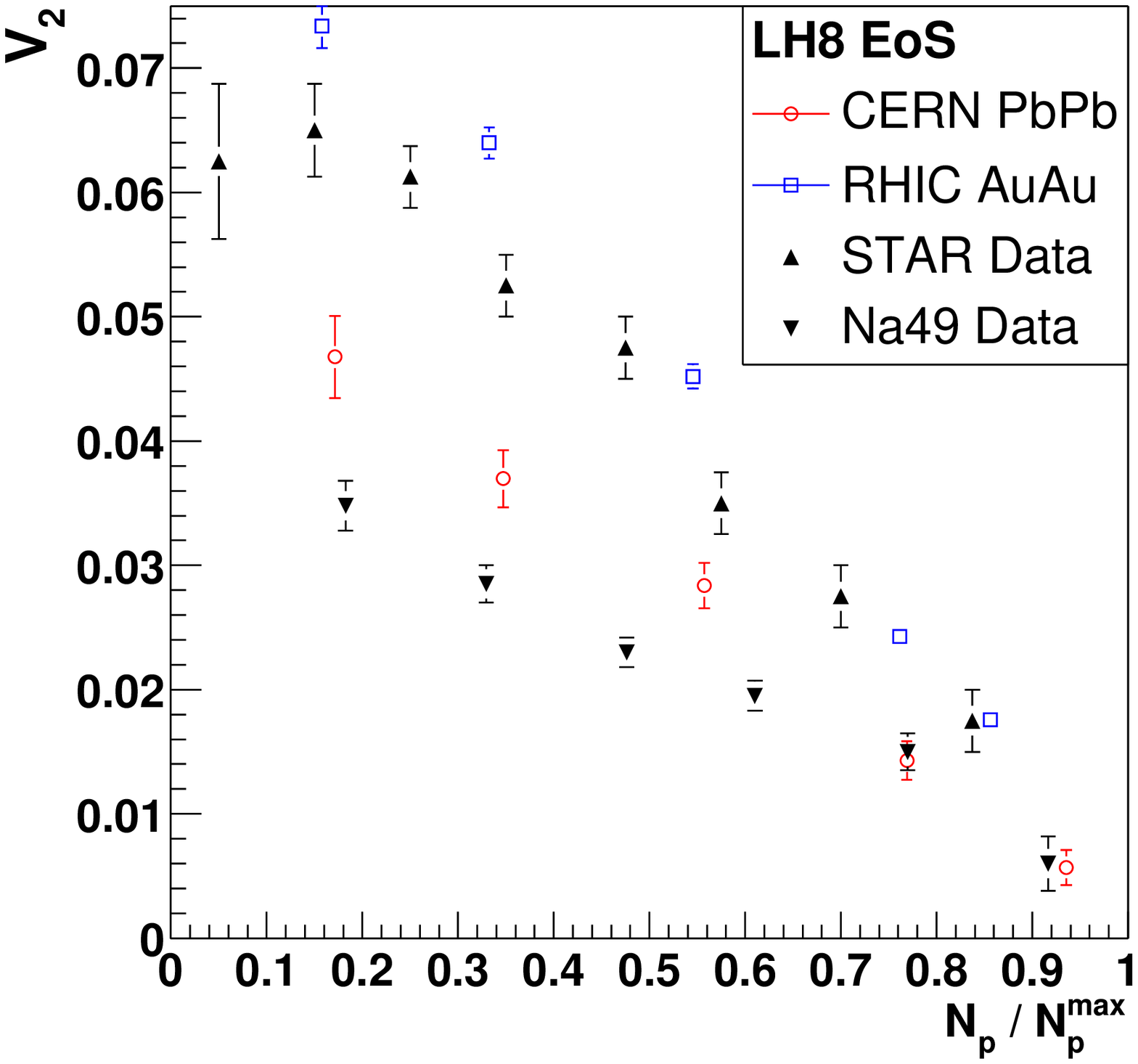} \hspace{-0.5cm}}
\vspace{-0.5cm}
\caption{\small From left: $v_2$ and $v_2(p_T)$ from hydrodynamics
\cite{KHHH00} and data from STAR \cite{v2_STAR}. On right: $v_2$ from 
hydrodynamics followed by RQMDv2.4 \cite{TLS00}.  }
\label{v2_hydro}
\vspace{-0.7cm}
\end{figure}

In conclusion, the first results from RHIC have again demonstrated the
importance of a close interplay between theory and experiments:
predictions for observables are difficult to obtain from truly first
principles. Indispensable constraints for the models are obtained from
certain sets of observables, such as the ones considered here. Especially
useful will be the studies of the systematics in $\sqrt s$, $A$ and {\bf b}.
When the production and evolution stage of the strongly interacting 
system are theoretically under control, the signals of the 
QGP can be predicted and found.

{\small 
\subsection*{Acknowledgements} 
\noindent I thank K. Tuominen, K. Kajantie, P. Huovinen,
N. Armesto, and V. Ruuskanen for help in preparing this talk,
V.J. Kolhinen for help with the eps-figures, X.-N. Wang for sending me  
Fig.~\ref{EKS98_fig}a, H. Honkanen for preparing
Fig.~\ref{EKS98_fig}c, and the Academy of Finland for financial
support.  }

\end{document}